\documentstyle[12pt]{ioplppt}
\def\journal#1,  #2,  #3   {
         {\sl #1~}{\bf #2}  #3    }

\def\pra{\journal Phys. Rev. A, }
\def\prb{\journal Phys. Rev. B, }
\def\prd{\journal Phys. Rev. D, }
\def\prl{\journal Phys. Rev. Lett., }

\def\npb{\journal Nucl. Phys. B, }

\def\plb{\journal Phys. Lett. B, }

\def\mpla{\journal Mod. Phys. Lett. A, }

\def\jp{\journal J. Phys. A: Math. Gen., }
\def\beq{\begin{equation}}
\def\eeq{\end{equation}}
\def\ba{\begin{eqnarray}}
\def\ea{\end{eqnarray}}
\def\nn{\nonumber \\}
\def\l{\lambda}

\def\f{\varphi}

\newcommand{\cl}{Calogero}
\newcommand{\st}{Sutherland}
\newcommand{\h}{Hamiltonian}
\newcommand{\col}{collective}
\newcommand{\pv}[1]{{-  \hspace {-4.0mm} #1}}
\newcommand{\rf}{\rho(\f)}

\begin{document}
\jl{1}
\title{Multi-vortex solution in the Sutherland model}
\author{I. Andri\'c, V. Bardek and L. Jonke\footnote{e-mail address: andric@thphys.irb.hr \\ \hspace*{3cm} bardek@thphys.irb.hr \\ \hspace*{3cm} larisa@thphys.irb.hr }}
\address{Theoretical Physics Division,\\
Rudjer Bo\v skovi\'c Institute, P.O. Box 1016,\\
10001 Zagreb, CROATIA}
\date{\today}
\begin{abstract}
We consider the large-$N$ Sutherland model in the Hamiltonian collective-field 
approach based on the $1/N$ expansion. The Bogomol'nyi limit appears and the  
corresponding solutions are given by static-soliton configurations. They exist
only for $\l<1$, i.e. for the negative coupling constant of the Sutherland 
interaction. We determine their creation energies and show that they are 
unaffected by higher-order corrections. For $\l=1$, the Sutherland model 
reduces to the free one-plaquette Kogut-Susskind model. 
\end{abstract}
\vspace{1cm}
\maketitle

\section{Introduction}

Classical and quantum solitons in the Calogero-Sutherland model have recently
been intensively studied \cite{camp,pol,mi}. The underlying motivation is 
that these extended objects may presumably play an important role in a deeper
understanding of quasiparticle and quasihole physics of exact solution 
 \cite{haa}. The \col \ theory 
of \cite{mi} offers a field-theoretic framework for describing 
semiclassical soliton configurations in the large-$N$ limit. We have recently 
shown that in the Calogero model there exist soliton-type finite-energy 
 solutions that can be obtained by solving a first-order integro-differential 
 equation of the Bogomol'nyi type.

In this paper we are primarily concerned with static solitons 
in the \col -field formulation of the Sutherland model. These solitons can 
also be reached by the Bogomol'nyi saturation. Among them there are some 
new, periodic multi-vortex solutions which have not been discussed so far.
 Their existence stems from the fact that the Sutherland model is 
defined on the compact support (circle) and therefore must satisfy periodic 
boundary conditions.
 In addition, the \col -field formulation of the $\l=1$ Sutherland \h \ 
transparently displays the equivalence to the one-plaquette restriction of the 
Kogut-Susskind $U(N)$ model \cite{kog}.

\section{The collective-field Hamiltonian}

The Sutherland system has a Hamiltonian describing spinless particles confined 
to a ring and interacting through a $1/r^2$ pairwise potential \cite{suth}:
\beq \label{h1}
H=\frac{4\pi^2}{L^2}\left( -\frac{1}{2}\sum_{i=1}^N\frac{d^2}{d\f_i^2}+
\frac{g}{4}\sum_{i>j}^N\frac{1}{\sin^2\left(\frac{\f_i-\f_j}{2} \right)}
\right) \eeq
where $L$ is the length of the ring and $N$ is the number of particles. 
We use units
in which $\hbar=m=1$, with $m$ being the mass of the particles.
Here $\f_i$ is the angular coordinate of the i-th particle. The dimensionless
 coupling constant $g$ determines the strength of the \st \ pair coupling 
and is related to the statistical parameter $\l$ of the exclusion statistics
 \cite{hal} by the relation \beq g=\l(\l-1) . \eeq
For special values of $\l$, i.e. $\l=0$, we have free bosons, and for $\l=1
\; {\rm and}\;2$, the model is related to a system of free fermions and to 
the Haldane-Shastry spin chain \cite{hssc}.
Because of the singularity of the \h \ for $\f_i=\f_j$, the wavefunction 
ought to have a prefactor that vanishes for coinciding particles:
\beq \label{DD} 
\psi=\Delta^{\l}\phi,\;\;\Delta=\prod_{i<j}\sin\left(\frac{\f_i-\f_j}{2}
\right) .\eeq
With this factorization, we obtain a new \h \ that acts on the residual, 
completely symmetric wavefunction $\phi$:
\beq \label{h2}
H=-\frac{1}{2}\sum_{i=1}^N\frac{d^2}{d\f_i^2}-\frac{\l}{2}\sum_{i=1}^N\left(
\sum_{j\neq i}^N\cot\frac{\f_i-\f_j}{2} \right)\frac{d}{d\f_i}+\frac{\l^2}
{24}N(N^2-1) .\eeq
The last constant term emerges from the trigonometric identity:
\beq \label{ti}
\sum_{i\neq j\neq k}\cot\frac{\f_i-\f_j}{2}\cot\frac{\f_i-\f_k}{2}=
-\frac{1}{3}N(N-1)(N-2). \eeq
The nontrivial part of the \st\  \h \ is now suitable for transformation into a 
\col-field representation. For the \col \ field we take the permutation 
symmetric function \beq \label{r1}
\rho(\f)=\sum_{i=1}^N\delta(\f-\f_i) \eeq obeying the normalization condition
\beq \label{r2} \int_0^{2\pi} d\f\rho(\f)=N .\eeq
Next, we reformulate the differential operators in the \h \ (\ref{h2}) in terms 
of a functional differentiation with respect to the \col \ field $\rho(\f)$.
For  $\l=1$, we have already written the \col -field version of the \h \ 
(\ref{h2}) in \cite{a2} and, for general $\l$, we proceed in a 
similar way.
Using the chain rule \beq \label{cr}
\frac{d}{d\f_i}=\int d\f \frac{\partial\rho(\f)}{\partial\f_i}\frac{\delta}
{\delta\rho(\f)} \eeq
and by rescaling the wavefunction 
\beq \phi(\f_1,...,\f_N)=J^{1/2}\Phi(\rho)  \eeq
after some calculation we find the Hermitian \col -field \h\
\ba \label{h3}
\fl H= \frac{1}{2}\int d\f\rf (\partial_{\f}\pi(\f))^2+
\frac{1}{8}\int d\f\rf\left(\partial_{\f}
\frac{\delta \ln J}{\delta\rf}\right)^2 \nn
\fl - \frac{\l-1}{4}\int d\f\partial_{\f}^2\delta(\f-\f')|_{\f=\f'} 
-\frac{\l}{4}\int d\f \rf\partial_{\f}\cot\frac{\f-\f'}{2}|_{\f=\f'}+
\frac{\l^2}{24}N(N^2-1). \ea
Here $\pi(\f)$ is the canonical conjugate of the field $\rf$:
\beq \label{com}
[\partial_{\f}\pi(\f),\rho(\f')]=-i\partial_{\f}\delta(\f-\f'). \eeq
The Jacobian $J$ is determined from the hermiticity condition
\beq \fl \partial_{\f}\left(\rf\partial_{\f}\frac{\delta \ln J}
{\delta\rf}\right)
=(\l-1)\partial_{\f}^2\rf+\l\partial_{\f}\left(\rf
\pv{\int} d\f'\cot\frac{\f-\f'}{2}\rho(\f')\right)  \eeq
and reads \beq \fl J=\exp\left[(\l-1)\int d\f \rf\ln\rf+\frac{\l}{2}
\int d\f d\f'\rf\ln\sin^2\frac{\f-\f'}{2}\rho(\f')\right] .\eeq
The two singular terms in the \h \ (\ref{h3}) do not contribute in the 
leading order of $N$. 
They should be cancelled by the infinite zero-point fluctuations of the \col \ 
field $\rf$. 
This will be discussed in detail in section 4.

\section{The Bogomol'nyi limit}

To find the ground-state energy and the corresponding \col \ 
motion in the large-$N$ limit, we should minimize the energy functional with 
respect to $\pi(\f)\;{\rm and}\;\rf$. However, in our case, owing to the 
special features of the model, there is a much more efficient method of 
minimization. The leading part of the \col -field \h \ in the $1/N$ 
expansion is given by the effective potential
\beq \label{v1}
\fl V_{\rm eff}=  \frac{1}{8}\int d\f\rf\left(\partial_{\f}\frac{\
\delta\ln J}{\delta\rf}\right)^2= 
 \frac{1}{2}\int d\f\rf\left(\frac{\l-1}{2}\frac{\partial_{\f}\rf}
{\rf}+\frac{\l}{2}\pv{\int} d\f'\cot\frac{\f-\f'}{2}\rho(\f')\right)^2 .\eeq
Owing to the positive definiteness of the effective potential (\ref{v1}), the 
Bogomol'nyi limit appears. The Bogomol'nyi bound is saturated by the positive, 
normalizable solution $\rho_0(\f)$ of the equation
\beq \label{b1}
\frac{\l-1}{2}\frac{\partial_{\f}\rf}{\rf}+\frac{\l}{2}\pv{\int}
d\f'\cot\frac{\f-\f'}{2}\rho(\f')=0 \eeq with the ground-state energy equal to
\beq \label{e0} E_{0}=\frac{\l^2}{24}N(N^2-1) \eeq which is the 
exact result \cite{suth}.
 The most obvious solution is given by the constant-density configuration
 $\rho=\rho_0$ for any value of the statistical parameter $\l$.
However, there exists one non-trivial solution to equation (\ref{b1}), 
given by \beq \label{val}
\rf=\frac{N}{2\pi}\frac{\sqrt{a^2-1}}{a+\cos n\f}  \eeq
where $a$ is an arbitrary positive parameter, $a>1$, and $n$ is an integer 
given by \beq \label{n} n=\frac{\l N}{\l-1} .\eeq
Indeed, by using
\numparts
\beq \int\frac{d\f}{a+\cos n\f}=\frac{2\pi}{\sqrt{a^2-1}} \eeq
\beq \pv{\int} d\f'\cot\frac{\f-\f'}{2}\frac{1}{a+\cos n\f'}=-\frac{2\pi}
{\sqrt{a^2-1}}\frac{\sin n\f}{a+\cos n\f} \eeq
\endnumparts
we can easily recover the solution (\ref{val}) and the constraint (\ref{n}). 
The solution (\ref{val}) exists only for special values of the 
statistical parameter, 
given by (\ref{n}). This
 constraint is a consequence of the periodicity condition
\beq \label{per} \rf=\rho(\f+2\pi). \eeq
It represents some kind of stationary waves around the constant condensed 
state $\rho_0=N/2\pi$.

Let us now find an interesting stationary hole-like excitation of 
 particles in the \st \ model, which can also be reached by the Bogomol'nyi 
saturation. Using the identity 
\ba \label{id}
&& {\cal P}\cot\frac{\f-\f'}{2}{\cal P}\cot\frac{\f-\f''}{2}+{\cal P}\cot
\frac{\f'-\f}{2}{\cal P}\cot\frac{\f'-\f''}{2} \nn
&& +{\cal P}\cot\frac{\f''-\f'}{2}{\cal P}\cot\frac{\f''-\f}{2}=4\pi^2
\delta(\f-\f')\delta(\f-\f'')-1 \ea
we can rewrite the \col -field potential $V_{\rm eff}$ as
\ba \label{v2}
\fl V_{\rm eff}=\frac{1}{2}\int d\f\rf\left(\frac{\l-1}{2}
\frac{\partial_{\f}\rf}
{\rf}+\frac{\l}{2}\pv{\int} d\f'\cot\frac{\f-\f'}{2}\rho(\f')+\frac{c}{2}\cot
\frac{\f}{2}\right)^2+\frac{c\l}{8}N^2 \nn
\fl +\frac{c-\l c-c^2}{8}\int d\f\rf\cot^2\frac{\f}{2}-\frac{\l-1}{8}cN
-\frac{c\l\pi^2}{2}\rho^2(0)+\frac{c\l}{8}
\left(\int d\f\rf\cot\frac{\f}{2}\right)^2 .\ea
For the symmetric configuration $\rf=\rho(-\f)$, representing a hole located at
the origin, $\rho(0)=0$, and for the particular 
value of the constant $c$ given by
\beq \label{c1} c=1-\l \eeq the Bogomol'nyi limit appears. The contribution 
of the squared term in $V_{\rm eff}$ vanishes and the corresponding 
configuration 
 satisfies the enlarged Bogomol'nyi equation
\beq \label{b2}
\frac{\l-1}{2}\frac{\partial_{\f}\rf}{\rf}+\frac{\l}{2}\pv{\int}
 d\f'\cot\frac{\f-\f'}{2}\rho(\f')+\frac{1-\l}{2}\cot\frac{\f}{2}=0.\eeq
The new, singular term in equation (\ref{b2}) is to compensate 
for the singularity produced by $\partial_{\f}\ln\rf$ at the origin, $\f=0$.
Equation (\ref{b2}) can be solved by a rational ansatz, and the normalized, 
static solution is of the form
\beq \label{rh}
\rf=\frac{a\sin^2\frac{\f}{2}}{b^2+\sin^2\frac{\f}{2}} \eeq where the constants 
$a$ and $b$ satisfy the constraint
\beq \label{const} -\frac{2\l}{\l-1}\frac{ab\pi}{\sqrt{1+b^2}}=1 .\eeq
>From (\ref{const}) and the normalization condition 
\beq \label{n2}
\int d\f\rho(\f)=N=2\pi a\left(1-\frac{b}{\sqrt{1+b^2}}\right) \eeq
it follows that 
\beq \label{cte}
a=\frac{N}{2\pi}+\frac{1-\l}{2\pi\l}\;
\;{\rm and}\;\;b^2=\frac{1}{ \left(N\frac{\l}{1-\l}+1\right)^2-1}. \eeq
Since the \col -field density is positive, $a$ and $b$ are necessarily positive
 parameters and therefore it follows from relation (\ref{const}) that $\l<1$.
The corresponding energy is given by \beq \label{e1}
E=E_0+\frac{N}{8}(1-\l)(\l N-\l+1) .\eeq 

We are now going to show that there exists a multi-vortex solution to 
equation (\ref{b2}). For this purpose, we must rearrange 
the effective potential 
$V_{\rm eff}$ as follows:
\ba \label{v3}
 V_{\rm eff}&&=\frac{1}{2}\int d\f\rf\left(\frac{\l-1}{2}\frac{\partial_{\f}
\rf}{\rf}+\frac{\l}{2}\pv{\int} d\f'\cot\frac{\f-\f'}{2}\rho(\f')+c\cot\frac
{n\f}{2}\right)^2 \nn &&-c\int 
d\f\rf\cot\frac{n\f}{2}\left(\frac{\l-1}{2}\frac
{\partial_{\f}\rf}{\rf}+\frac{\l}{2}
\pv{\int} d\f'\cot\frac{\f-\f'}{2}\rho(\f')\right) \nn
&&-\frac{c^2}{2}\int d\f\rf\cot^2\frac{n\f}{2} . \ea 
In order to get the Bogomol'nyi form, we should show that all terms in 
$V_{\rm eff}$, except the first one, transform into an irrelevant constant 
functional.
All $\int d\f\rf\cot^2\frac{n\f}{2}$ terms disappear if the  strength 
$c$ of the 
cotangens regulator term is given by \beq \label{c2} c=\frac{1-\l}{2}n .\eeq
Using the summation formula \cite{rus}  \beq \label{f}
n\cot\frac{n\f}{2}=\sum_{k=0}^{n-1}\cot\left(\frac{\f}{2}+\frac{k\pi}{n}\right)
 \eeq and the principal-value identity (\ref{id}), we can recast the final 
term in $V_{\rm eff}$ as \ba \label{v4}
&& n\frac{\l(\l-1)}{4}\int d\f d\f'\rf\rho(\f')\cot\frac{\f-\f'}{2}
\cot\frac{n\f}{2} \nn &&=\frac{\l(\l-1)}{4}\left[-n\frac{N^2}{2}+2\pi^2
\sum_{k=0}^{n-1}\rho^2\left(\frac{2k\pi}{n}\right)\right] .\ea
Here we have assumed that $\rf$ is an even function in $\f$. Assuming further 
that $\rf$ describes the n-vortex-like configuration, with equidistant 
vanishing points at $\f_k=2k\pi/n$, our $V_{\rm eff}$ functional finally 
reduces to \ba \label{v5}
\fl V_{\rm eff}=\frac{(\l-1)^2}{8}Nn^2-\frac{\l(\l-1)}{8}N^2n \nn
\fl + \frac{1}{2}\int d\f\rf\left(\frac{\l-1}{2}\frac{\partial_{\f}
\rf}{\rf}+\frac{\l}{2}\pv{\int} d\f'\cot\frac{\f-\f'}{2}\rho(\f')+n\frac{1-\l}
{2}\cot\frac{n\f}{2}\right)^2 .\ea
We have achieved our goal and the minimal value of $V_{\rm eff}$ is given 
by the Bogomol'nyi saturation:
\beq \label{b3}
\frac{\l-1}{2}\frac{\partial_{\f}\rf}{\rf}+\frac{\l}{2}\pv{\int}
 d\f'\cot\frac{\f-\f'}{2}\rho(\f')+n\frac{1-\l}{2}\cot\frac{n\f}{2}=0 .\eeq
The corresponding energy is given by \beq \label{en}
E_n=E_0+\frac{(\l-1)^2}{8}Nn^2-\frac{\l(\l-1)}{8}N^2n .\eeq
For $n=1$, the energy is equal to that found in (\ref{e1}). The enlarged 
Bogomol'nyi equation can again be solved by a rational ansatz
\beq \label{rh2}
\rf=\frac{p\sin^2\frac{n\f}{2}}{q^2+\sin^2\frac{n\f}{2}}. \eeq
Contour integration gives
\beq \label{ci}
\pv{\int}_{-\pi}^{\pi} d\f'\cot\frac{\f-\f'}{2}\rho(\f')=-\frac{pq\pi}{\sqrt{1+q
^2}}\frac{\sin n\f}{q^2+\sin^2\frac{n\f}{2}}. \eeq
Substituting equation (\ref{ci}) into (\ref{b3}), we obtain the 
following condition for the positive parameters 
$p$ and $q$: \beq \label{par}
-\frac{2\l}{\l-1}\frac{pq\pi}{\sqrt{1+q^2}}=n .\eeq
>From relation (\ref{par}) and the normalization condition equivalent
 to (\ref{n2}) we find that the parameters 
$p$ and $q$ read 
\numparts
\beq p=\frac{1}{2\pi}\left( N+n\frac{1-\l}{\l}\right) \eeq
\beq q^2=\frac{1}{\left(\frac{\l}{1-\l}\frac{N}{n}+1\right)^2-1} .\eeq
\endnumparts
It is evident that the constraint (\ref{par}) implies $\l<1$ .

\section{Quantum corrections}

Let us now turn our attention to the next-to-leading-order terms in 
the \col \ \h \ (\ref{h3}). We are going to study the effect of the small 
density fluctuations around  the hole-like configuration:
\beq  \rf=\rho_0(\f)+\eta(\f) .\eeq
Introducing the operators
\numparts
\beq \label{a1}
\fl A(\f)=\partial_{\f}\pi(\f)+i\left[\frac{\l -1}{2}\partial_{\f}\left
(\frac{\eta(\f)}
{\rho_0(\f)}\right)+\frac{\l}{2}\pv{\int} d\f'\cot\frac{\f-\f'}{2}\eta(\f')
-\frac{\l -1}{2}\cot\frac{\f}{2}\right]  \eeq 
\beq \label{a2}
\fl A^{\dagger}(\f)=\partial_{\f}\pi(\f)-i\left[\frac{\l -1}{2}
\partial_{\f}\left(
\frac{\eta(\f)}{\rho_0(\f)}\right)+\frac{\l}{2}\pv{\int} d\f'\cot\frac{
\f-\f'}{2}\eta(\f')-\frac{\l -1}{2}\cot\frac{\f}{2}\right]  \eeq
\endnumparts
with the c-number commutator
\beq \label{comm}
[A(\f),A^{\dagger}(\f')]=-(\l-1)\partial_{\f}\partial_{\f'}\frac{\delta(
\f-\f')}{\rho_0(\f)}+\l\partial_{\f}\cot\frac{\f-\f'}{2}  \eeq
the \col \ \h \ can be written up to the quadratic terms in $\eta$ and $\pi$ as
\beq \label{nH}
H=E_0+\frac{N}{8}(1-\l)(\l N-\l+1)+\frac{1}{2}\int d\f
{\rho_0(\f)}A^{\dagger}(\f)A(\f) .\eeq
The divergent terms disappear, as can be easily checked using the commutator 
(\ref{comm}). The \col \ \h \ is semidefinite and there exists the \col -field 
wavefunctional $\Phi(\eta)$ such that \beq \label{**}
A(\f)\Phi(\eta)=0 .\eeq
For this wavefunctional the correction due to fluctuations is vanishing.
Solving equation (\ref{**}), we easily get 
\ba \label{phi}
\Phi(\eta)=\exp && \left\{\frac{\l-1}{4}\int d\f\frac{\eta^2(\f)}{\rho_0(\f)}
+\frac{\l}{4}\int d\f d\f'\eta(\f)\ln\sin^2\frac{\f-\f'}{2}\eta(\f) 
\right. \nn
 && \left. -\frac{\l-1}{2}\int d\f\ln\sin^2\frac{\f}{2}\eta(\f)\right\} .\ea
>From this result we can reconstruct the Schr\"odinger wavefunction
$\Psi(\f_1,...,\f_N)$ for the $N$-particle system, which corresponds to the 
one-hole configuration. It is given by
\beq \label{sch}
\Psi(\eta)=\Delta^{\l} J^{1/2}\Phi(\eta) .\eeq
Here, the $\Delta $ prefactor is present owing to the extraction (\ref{DD}).
The Jacobian of the transformation from $\f_i$ into $\rf$ rescales the 
wavefunctional by the $ J^{1/2}$ factor. By expanding the Jacobian to the 
quadratic terms in $\eta$ and using the relations (\ref{phi}), (\ref{sch}) 
and the Bogomol'ny equation for $\rho_0$ (\ref{b2}), we are left with 
\beq \label{sch2}
\Psi(\eta)=\Delta^{\l}\exp\left[\frac{1-\l}{2}\int d\f\ln\sin^2
\frac{\f}{2}\eta(\f)\right] 
 .\eeq  If we substitute equation (\ref{r1}) into (\ref{sch2}), we obtain the
 wavefunction for $N$ particles
\beq \label{sch3}
\psi(\f_1,...,\f_N)=\Delta^{\l}\prod_{i=1}^N\sin^{1-\l}\frac{\f_i}{2} .\eeq
It can be easily checked that this wave function indeed describes the 
configuration with the known energy (\ref{e1}), provided that $\rf$ satisfies
\beq \label{rhoo}
\int d\f\cot\frac{\f}{2}\rf =0,\;\;\; {\rm i.e.},\;\;\;
\sum_{i=1}^N\cot\frac{\f_i}{2}=0 . \eeq
The wavefunctional for the n-hole-like configuration can be formed along 
similar lines, explicitly given for the one-hole case. It reads
\beq \label{schN}
\psi_n(\f_1,...,\f_N)=\Delta^{\l}\prod_{i=1}^N\sin^{n(1-\l)}
\frac{n\f_i}{2} .\eeq 
It can be shown that this wavefunction is indeed the eigenfunction of the 
$N$-particle Hamiltonian (\ref{h1}) with the known energy 
(\ref{en}), provided that $\rf $ satisfies 
\beq \label{ner}
\int d\f \cot\frac{n\f}{2}\rf =0,\;\;\; {\rm i.e.},\;\;\;
\sum_{i=1}^N\cot\frac{n\f_i}{2}=0 . \eeq

Let as briefly comment on the corresponding quantum corrections in the case 
of the Calogero model. Owing to the Bogomol'nyi form we can in the same 
way show the stability of the static solitons \cite{mi} against first-order 
quantum corrections.

\section{Equivalence with the one-plaquette $U(N)$ gauge theory }

Finally, let us show that there is equivalence of the $\l=1$ Sutherland model 
and the free one-plaquette Kogut-Susskind lattice gauge theory. 
The one-plaquette restriction of the $U(N)$ lattice gauge theory in $2+1$ 
dimensions is given by 
\beq \label{51}
\fl H=\frac{g^2}{2a}\left\{\sum_{\alpha,i=1}^4 E^{\alpha}(i)E^{\alpha}(i)+
\frac{2}{g^4}S[U(1)U(2)U(3)U(4)]\right\}\;\;\alpha=0,1,...,N^2-1 \eeq
where $g$ is the coupling constant and $a$ is the lattice spacing. 
The basic degrees of freedom are given by the unitary matrices $U(i)$, 
whereas the electric field $E^{\alpha}(i)$ represents the conjugate variable 
in the  vertex $i$ of the plaquette. The lattice action $S$ is given by the 
real function on the group $U(N)$. In the \col -field method, we rewritte the 
\h \ in terms of Wilson loop variables
\beq \label{wil}
W_n={\rm Tr}\left\{[U(1)U(2)U(3)U(4)]^n\right\} \eeq
where $n$ is an integer, or its continuous version
\beq \label{wcont}
\rf =\sum_{n=-\infty}^{+\infty} \frac{1}{2\pi}e^{in\f} W_n .\eeq
Here we simply quote the final results and refer the reader to \cite{a2}
 for their derivation. It turns out that the \col -field version of 
the \h \ (\ref{51}) is 
\ba \label{c51}
H &&= \frac{2g^2}{a}\pv\int d\f\rf\left[(\partial_{\f}\pi)^2+
\frac{1}{8}\left(\pv\int d\f'\rho(\f')\cot\frac{\f-\f'}{2}\right)^2\right] \nn
&& -\frac{g^2}{2a}\pv\int d\f\rf\partial_{\f}\cot\frac{\f-\f'}{2}|_{\f=\f'}+
\frac{1}{g^2a}\int d\f\rf S(\f) .\ea
So, apart from an overall constant $2g^2/a$, and the last interaction term, 
the \h \ (\ref{c51}) is identical to the Sutherland \h \ (\ref{h3}) for 
fermions, i. e. for $\l=1$. Moreover, using the identity (\ref{id}) we 
recover the \col -field \h \ for the $c=1$ matrix model, up to the 
irrelevant constant term \cite{antal}:
\beq \label{cm51}
\fl H=\frac{4g^2}{a}\left\{\frac{1}{2}\int d\f\rf(\partial_{\f}\pi)^2+
\frac{\pi^2}{6}\int d\f\rho^3(\f)-\frac{N^3}{24}+\frac{1}{4g^4}
\int d\f\rf S(\f)\right\} \eeq
the only difference being that fermions live on a circle and interact with 
an external potential $\frac{1}{4g^4}S(\f)$.

\section{Summary}

We have found three main results. The first result is that  
in the  \col -field formulation of the Sutherland model
 there exist multi-vortex 
static configurations which could be reached by the  Bogomol'nyi saturation. 
Since we know that in the Calogero model there exists a moving 
soliton \cite{pol,mi},
it is of interest to look for the existence of moving-multi-vortex solutions.

The second result is that the energies of these configurations 
 are not affected by the next-to-leading-order  
 corrections stemming from the quantum \col -field fluctuations.
This is what happens in the supersymmetric theory where the Bogomol'nyi bound 
does not receive quantum corrections \cite{w}. Therefore we conclude that there 
must exist a supersymmetric extension of the \cl -Sutherland model in the 
\col -field formulation. For $\l=1$, this has already been done in \cite{J}.

The third result is that the $2+1$-dimensional gluodynamics with the 
$U(N)$ gauge 
group is, in the large-$N$ limit, equivalent to the system of $N$ 
non-relativistic fermions on a circle. There is some resemblance with the 
results of \cite{min}, where the equivalence of $1+1$-dimensional 
 QCD and the $c=1$ matrix model was found. However, in our case, 
the dimension is higher and  fermions are not free, but interact with 
some sort of external potential originating from the corresponding
one-plaquette action.

\ack

This work was supported by the Scientific Fund of the Republic of Croatia.


\end{document}